\documentclass{revtex4}
\usepackage{amssymb}
\usepackage{amsfonts}
\usepackage{amsmath}
\usepackage{graphicx}
\usepackage{float}

\begin{document}

\title{PDM-charged particles in PD-magnetic plus Aharonov-Bohm flux fields:
unconfined "almost-quasi-free" and confined in a Yukawa plus Kratzer exact
solvability}
\author{Omar Mustafa }
\email{omar.mustafa@emu.edu.tr}
\author{Zeinab Algadhi}
\email{zeinab.algadhi@emu.edu.tr}
\affiliation{Department of Physics, Eastern Mediterranean University, G. Magusa, north
Cyprus, Mersin 10 - Turkey,\\
Tel.: +90 392 6301378; fax: +90 3692 365 1604.}

\begin{abstract}
\textbf{Abstract:} Using azimuthally symmetrized cylindrical coordinates, we
consider some position-dependent mass (PDM) charged particles moving in
position-dependent (PD) magnetic and Aharonov-Bohm flux fields. We focus our
attention on PDM-charged particles with $m\left( \overrightarrow{r}\right)
=g\left( \rho \right) =\eta \,f\left( \rho \right) \exp \left( -\delta \rho
\right) $ (i.e., the PDM is only radially dependent) moving in an inverse
power-law-type radial PD-magnetic fields $\overrightarrow{B}=B_{\circ
}\left( \mu /\rho ^{\sigma }\right) \widehat{z}$. Under such settings, we
consider two \emph{almost-quasi-free} PDM-charged particles (i.e., no
interaction potential, $V\left( \overrightarrow{r}\right) =0$) endowed with $%
g\left( \rho \right) =\eta /\rho $ and $g\left( \rho \right) =\eta /\rho
^{2} $. Both yield exactly solvable Schr\"{o}dinger equations of Coulombic
nature but with different spectroscopic structures. Moreover, we consider a
Yukawa-type PDM-charged particle with $g\left( \rho \right) =\eta \,\exp
\left( -\delta \rho \right) /\rho $ moving not only in the vicinity of the
PD-magnetic and Aharonov-Bohm flux fields but also in the vicinity of a
Yukawa plus a Kratzer type potential force field $V\left( \rho \right)
=-V_{\circ }\exp \left( -\delta \rho \right) /\rho -V_{_{1}}/\rho
+V_{_{2}}/\rho ^{2}$. For this particular case, we use the Nikiforov-Uvarov
(NU) method to come out with exact analytical eigenvalues and
eigenfunctions. Which, in turn, recover those of the \emph{almost-quasi-free}
PDM-charged particle with $g\left( \rho \right) =\eta /\rho $ for $V_{\circ
}=V_{_{1}}=V_{_{2}}=0=\delta $. Energy levels crossings are also reported.


\end{abstract}

\maketitle

\section{Introduction}

A charged particle moving in a uniform/constant magnetic field and/or an
Aharonov-Bohm flux field has been a subject of research interest over the
years \cite{1,2,3,4,5,6,7,8,9,10}. On the classical mechanical and
mathematical side of the problem, it is crucial to know that the canonical
momentum is no longer the mass times velocity but an extra term is added so
that $p_{_{i}}=m_{\circ }v_{_{i}}+eA_{_{i}}$ (where $m_{\circ }$ is the
conventional constant mass, $e$ is the charge of the particle and $A_{_{i}}$
is the $i$th component of the vector potential). The problem is readily of a
delicate nature, especially when the magnetic field is no longer a constant
but rather a position-dependent one (to be referred to as PD-magnetic field,
hereinafter). Moreover, particles endowed with position-dependent mass (PDM)
are considered interesting and unavoidable in both quantum and classical
mechanics \cite%
{11,12,13,14,15,16,17,18,19,20,21,22,23,24,25,26,27,28,29,30,31,32,33,34,35,36}%
. Such mass settings find their applications in condensed matter physics
(see, e.g., \cite{20,26,27,30}), in optical physics (see, e.g., \cite{37,38}%
), etc. They are not to be necessarily understood as particles with PDM
literally. A position-dependent deformation in the coordinate system may
very well render the mass position-dependent. One would then express the
mass as $M\left( \overrightarrow{r}\right) =m_{\circ }m\left( 
\overrightarrow{r}\right) $, where $m\left( \overrightarrow{r}\right) $ is a
dimensionless position-dependent scalar multiplier. It would be interesting,
therefore, to consider a PDM-charged particle moving in not only a
PD-magnetic field and an Aharonov-Bohm flux field but also in the vicinity
of a Yukawa-type plus a Kratzer-type molecular interaction force fields.
Hereby, we need to use the Nikiforov-Uvarov (NU) method (see e.g. \cite%
{39,40,41}) and explore its exact solvability. This forms a constituent
inspiration of the current methodical proposal.

A priori, we recollect that while in classical mechanics the PDM particles
cause no conflict at all (see, e.g., \cite{31,36}), they yield an ordering
ambiguity problem in quantum mechanics. This ambiguity is a manifestation of
the non-unique representation of the PDM kinetic energy operator (see e.g., 
\cite{12,13,22,24,36}) of the von Roos Hamiltonian \cite{12}. However, in
his analysis on the transition from classical PDM-Hamiltonians into quantum
mechanical PDM-Hamiltonians, Mustafa \cite{36} has argued (using $\hbar
=2m_{\circ }=1$ units) that whilst it is safe in classical mechanics to
write the PDM-kinetic energy term as%
\begin{equation}
T=\frac{\overrightarrow{P}\left( \overrightarrow{r}\right) ^{2}}{2m\left( 
\overrightarrow{r}\right) };\text{ \ }\overrightarrow{P}\left( 
\overrightarrow{r}\right) =m\left( \overrightarrow{r}\right) \vec{v},
\end{equation}%
it is necessary and convenient to write the quantum mechanical kinetic
energy operator as%
\begin{equation}
\widehat{T}=\left( \frac{\widehat{P}\left( \overrightarrow{r}\right) }{\sqrt{%
m\left( \overrightarrow{r}\right) }}\right) ^{2}.
\end{equation}%
Where $\widehat{P}\left( \overrightarrow{r}\right) $ is the PDM-momentum
operator. Consequently, the PDM-minimal coupling $\widehat{P}\left( 
\overrightarrow{r}\right) \longrightarrow \widehat{P}\left( \overrightarrow{r%
}\right) -e\overrightarrow{A}\left( \overrightarrow{r}\right) $ should be
indulged into the PDM-Schr\"{o}dinger equation \ as%
\begin{equation}
\left[ \left( \frac{\widehat{P}\left( \overrightarrow{r}\right) -e%
\overrightarrow{A}\left( \overrightarrow{r}\right) }{\sqrt{m\left( 
\overrightarrow{r}\right) }}\right) ^{2}+W\left( \overrightarrow{r}\right) %
\right] \psi \left( \overrightarrow{r}\right) =E\psi \left( \overrightarrow{r%
}\right) ;\ \ W\left( \overrightarrow{r}\right) =e\varphi \left( 
\overrightarrow{r}\right) +V\left( \overrightarrow{r}\right) ,
\end{equation}%
where $\overrightarrow{A}\left( \overrightarrow{r}\right) $ is the vector
potential, $e\varphi \left( \overrightarrow{r}\right) $ is a scalar
potential and $V\left( \overrightarrow{r}\right) $ is any other potential
energy than the electromagnetic one. In a subsequent work, moreover, Mustafa
and Algadhi \cite{11} have constructed and defined the PDM-momentum operator
as 
\begin{equation}
\widehat{P}\left( \overrightarrow{r}\right) =-i\left[ \overrightarrow{\nabla 
}-\frac{1}{4}\left( \frac{\overrightarrow{\nabla }m\left( \overrightarrow{r}%
\right) }{m\left( \overrightarrow{r}\right) }\right) \right] .
\end{equation}%
This would allow us to re-write equation (3) as%
\begin{eqnarray}
&&\left[ -\frac{1}{m\left( \overrightarrow{r}\right) }\overrightarrow{\nabla 
}^{2}+\left( \frac{\overrightarrow{\nabla }m\left( \overrightarrow{r}\right) 
}{m\left( \overrightarrow{r}\right) ^{2}}\right) \cdot \overrightarrow{%
\nabla }+\frac{1}{4}\left( \frac{\overrightarrow{\nabla }^{2}m\left( 
\overrightarrow{r}\right) }{m\left( \overrightarrow{r}\right) ^{2}}\right) -%
\frac{7}{16}\left( \frac{\left[ \overrightarrow{\nabla }m\left( 
\overrightarrow{r}\right) \right] ^{2}}{m\left( \overrightarrow{r}\right)
^{3}}\right) +\frac{2\ i\ e}{m\left( \overrightarrow{r}\right) }%
\overrightarrow{A}\left( \overrightarrow{r}\right) \cdot \overrightarrow{%
\nabla }\right.  \notag \\
&&\quad \quad \left. +\frac{ie}{m\left( \overrightarrow{r}\right) }\left( 
\overrightarrow{\nabla \,}\cdot \overrightarrow{A}\left( \overrightarrow{r}%
\right) \right) -i\ e\ \overrightarrow{A}\left( \overrightarrow{r}\right)
\cdot \left( \frac{\overrightarrow{\nabla }m\left( \overrightarrow{r}\right) 
}{m\left( \overrightarrow{r}\right) ^{2}}\right) +\frac{e^{2}\overrightarrow{%
A}\left( \overrightarrow{r}\right) ^{2}}{m\left( \overrightarrow{r}\right) }%
+W\left( \overrightarrow{r}\right) \right] \psi \left( \overrightarrow{r}%
\right) =E\psi \left( \overrightarrow{r}\right) ,
\end{eqnarray}%
in which the vector potential takes a conventional form that satisfies the
Coulomb gauge $\overrightarrow{\nabla }\cdot \overrightarrow{A}\left( 
\overrightarrow{r}\right) =0$ and results in a uniform constant magnetic
field through the traditional textbook recipe $\overrightarrow{\nabla }%
\times \overrightarrow{A}\left( \overrightarrow{r}\right) =\overrightarrow{B}%
=B_{\circ }\widehat{z}$. This magnetic field setting is the commonly and
frequently used in the literature (see e.g. \cite{42} and related references
cited therein). Nevertheless, in the construction of the vector potential $%
\overrightarrow{A}\left( \overrightarrow{r}\right) $, the magnetic field may
turn out to be a PD-magnetic field (see e.g., \cite{11,43}). In our current
proposal, we focus our attention on PDM-charged particles in PD-magnetic and
Aharonov-Bohm flux fields, without the confinement potential (i.e., $V\left( 
\overrightarrow{r}\right) =0$) and with a confinement potential (i.e., $%
V\left( \overrightarrow{r}\right) \neq 0$). The organization of this paper
is, therefore, in order.

In section 2, we consider a PDM-charged particle in PD-magnetic and
Aharonov-Bohm flux fields and discuss the separability of the PDM-Schr\"{o}%
dinger equation (5) using azimuthally symmetrized cylindrical coordinates $%
\left( \rho ,\varphi ,z\right) $. Therein, we use a general form of the
vector potential $\overrightarrow{A}\left( \overrightarrow{r}\right) $ so
that a radial PD-magnetic field emerges in the process (i.e., $%
\overrightarrow{B}=B_{\circ }F\left( \rho \right) \widehat{z}$, where $%
F\left( \rho \right) $ is a dimensionless radial scalar multiplier to be
discussed/determined below). In the same section, moreover, we construct our
PD-magnetic field in such a way that it is of a feasibly experimentally
applicable nature (i.e., inverse power-law type $\overrightarrow{B}=B_{\circ
}\left( \mu /\rho ^{\sigma }\right) \widehat{z}$) to be used along with a
PDM $m\left( \overrightarrow{r}\right) =g\left( \rho \right) =\eta \,f\left(
\rho \right) \exp \left( -\delta \rho \right) $ (i.e., the PDM is only
radial-dependent). In section 3, we consider the what may be called \emph{%
almost-quasi-free} PDM-charged particles (i.e., no other interaction
potential than the interaction of the PDM-charged particles with the
PD-magnetic and Aharonov-Bohm flux fields, where the conventional
confinement $V\left( \overrightarrow{r}\right) =0$) endowed with two
unavoidable exactly solvable PDM models $g\left( \rho \right) =\eta /\rho $
and $g\left( \rho \right) =\eta /\rho ^{2}$. A PDM-charged particle, with $%
m\left( \overrightarrow{r}\right) =g\left( \rho \right) =\eta \,\exp \left(
-\delta \rho \right) /\rho $; $f\left( \rho \right) =1/\rho $, interacting
with a PD-magnetic plus Aharonov-Bohm flux fields and moving in the vicinity
of a Yukawa plus a Kratzer type potential force field $V\left( \rho \right)
=-V_{\circ }\exp \left( -\delta \rho \right) /\rho -V_{_{1}}/\rho
+V_{_{2}}/\rho ^{2}$ is considered in section 4. Where the Nikiforov-Uvarov
(NU) method is used to obtain exact eigenvalue and eigenfunctions. The
potency of this method in obtaining exact analytical solutions is well
documented in the sample of references (see, e.g., \cite{6,39,40,41}). Such
exact results collapse into those of the \emph{almost-quasi-free}
PDM-charged particles with $m\left( \overrightarrow{r}\right) =g\left( \rho
\right) =\eta /\rho $, of section 3, when $V_{\circ
}=V_{_{1}}=V_{_{2}}=0=\delta $ are used (this should be the natural tendency
of the more general case, of course). The \emph{almost-quasi-free}
PDM-charged particles with $m\left( \overrightarrow{r}\right) =g\left( \rho
\right) =\eta /\rho $ would, therefore, play the role of an
exact-solvability test, so to speak. For the sample examples mentioned
above, we have studied the effects of all parametric settings involved in
the PDM, PD-magnetic field, and/or interaction potential on the spectra. We
have observed that energy levels crossings (that may very well be considered
as \emph{occasional degeneracies }at some specific parametric settings) are
unavoidable in the process. Such energy levels crossings are the signature
of the PDM settings. We discuss this issue in the concluding remarks section
5.

\section{PDM-charged particles in PD-magnetic and Aharonov-Bohm flux fields}

Let us start with PDM-Schr\"{o}dinger equation (5) and discuss its
separability under azimuthal symmetrization within the cylindrical
coordinates $\left( \rho ,\varphi ,z\right) $. Moreover, our PDM-charged
particle is of charge $e=\pm \left\vert e\right\vert $ and is considered to
be interacting with the vector potential%
\begin{equation}
\overrightarrow{A}\left( \overrightarrow{r}\right) =\overrightarrow{A}%
_{1}\left( \overrightarrow{r}\right) +\overrightarrow{A_{2}}\left( 
\overrightarrow{r}\right) ;\left\{ 
\begin{tabular}{l}
$\overrightarrow{A}_{1}\left( \overrightarrow{r}\right) =\left( 0,B_{\circ
}\rho \,S\left( \rho \right) /2,0\right) \medskip $ \\ 
$\overrightarrow{A}_{2}\left( \overrightarrow{r}\right) =\left( 0,\Phi
_{AB}/2\pi \rho ,0\right) $ \medskip%
\end{tabular}%
\right. ,
\end{equation}%
where a PD-magnetic field is manifested by the vector potential $%
\overrightarrow{A}_{1}\left( \overrightarrow{r}\right) $ so that%
\begin{equation}
\overrightarrow{B}=\overrightarrow{\nabla }\times \overrightarrow{A}%
_{1}\left( \overrightarrow{r}\right) =B_{\circ }\left[ S\left( \rho \right) +%
\frac{\rho }{2}S^{\prime }\left( \rho \right) \right] \widehat{z}\text{ };%
\text{ \ }S^{\prime }\left( \rho \right) =\frac{dS\left( \rho \right) }{%
d\rho }.
\end{equation}%
Here, $\overrightarrow{\nabla }\times \overrightarrow{A}_{2}\left( 
\overrightarrow{r}\right) =0.$ with $\overrightarrow{A}_{2}\left( 
\overrightarrow{r}\right) $ describing the Aharonov-Bohm flux field $\Phi
_{AB}$ effect (see, e.g., \cite{6,41,42}), and $S\left( \rho \right) $ is a
dimensionless scalar multiplier and is a byproduct of the construction
process of the vector potential $\overrightarrow{A}_{1}\left( 
\overrightarrow{r}\right) $ (note that the case $S\left( \rho \right) =1$
recovers the constant magnetic field settings). Consequently, our
PDM-charged particle interacts with the total vector potential

\begin{equation}
\overrightarrow{A}\left( \overrightarrow{r}\right) =\left( 0,\frac{B_{\circ }%
}{2}\rho \,S\left( \rho \right) +\frac{\Phi _{AB}}{2\pi \rho },0\right)
=\left( 0,A_{\varphi },0\right) .
\end{equation}

At this point, we use the assumptions that the PDM function is only radially
dependent, i.e.,

\begin{equation}
m\left( \overrightarrow{r}\right) =m\left( \rho ,\varphi ,z\right)
=\,g\left( \rho \right) ,
\end{equation}%
and $V\left( \varphi \right) =0$ to secure azimuthal symmetrization so that

\begin{equation}
g\left( \rho \right) W\left( \rho ,\varphi ,z\right) =V\left( \rho \right)
+V\left( z\right) .
\end{equation}%
This would, in turn, facilitate separability of the PDM-Schr\"{o}dinger
equation (5) at hand and allow the substitution of the wavefunction%
\begin{equation}
\psi \left( \overrightarrow{r}\right) =\psi \left( \rho ,\varphi ,z\right)
=R\left( \rho \right) Z\left( z\right) e^{im\varphi },
\end{equation}%
(where $m=0,\pm 1,\pm 2,...,\pm \ell $ \ is the magnetic quantum number, and 
$\ell $ is angular momentum quantum number) to obtain.

\begin{gather}
\frac{R^{\prime \prime }\left( \rho \right) }{R\left( \rho \right) }-\left( 
\frac{g^{\prime }\left( \rho \right) }{g\left( \rho \right) }-\frac{1}{\rho }%
\right) \frac{R^{\prime }\left( \rho \right) }{R\left( \rho \right) }-\frac{1%
}{4}\left( \frac{g^{\prime \prime }\left( \rho \right) }{g\left( \rho
\right) }+\frac{g^{\prime }\left( \rho \right) }{\rho g\left( \rho \right) }%
\right) +\frac{7}{16}\left( \frac{g^{\prime }\left( \rho \right) }{g\left(
\rho \right) }\right) ^{2}  \notag \\
-\frac{m^{2}}{\rho ^{2}}+\frac{2em}{\rho }A_{\varphi }-e^{2}A_{\varphi
}^{2}+g\left( \rho \right) E-V\left( \rho \right) -k_{z}^{2}=0.
\end{gather}%
Where, $k_{z}^{2}$\ represents the eigenvalues of the $z$-dependent part

\begin{equation}
\frac{Z^{\prime \prime }\left( z\right) }{Z\left( z\right) }-V\left(
z\right) -k_{z}^{2}=0.
\end{equation}%
Consequently, the radially-dependent part along with (8) reads

\begin{gather}
\left[ \frac{R^{\prime \prime }\left( \rho \right) }{R\left( \rho \right) }%
-\left( \frac{g^{\prime }\left( \rho \right) }{g\left( \rho \right) }-\frac{1%
}{\rho }\right) \frac{R^{\prime }\left( \rho \right) }{R\left( \rho \right) }%
-\frac{1}{4}\left( \frac{g^{\prime \prime }\left( \rho \right) }{g\left(
\rho \right) }+\frac{g^{\prime }\left( \rho \right) }{\rho g\left( \rho
\right) }\right) +\frac{7}{16}\left( \frac{g^{\prime }\left( \rho \right) }{%
g\left( \rho \right) }\right) ^{2}\right.  \notag \\
\left. -\frac{\tilde{m}^{2}}{\rho ^{2}}+e\tilde{m}B_{\circ }S\left( \rho
\right) -k_{z}^{2}-\frac{e^{2}B_{\circ }^{2}}{4}\left[ \rho S\left( \rho
\right) \right] ^{2}+g\left( \rho \right) E-V\left( \rho \right) \right] =0.
\end{gather}%
Here, $\alpha =\Phi _{AB}/\Phi _{\circ };$ $\Phi _{\circ }=2\pi /e,$ is the
Aharonov-Bohm flux quantum (within the current units, $\hbar =2m_{\circ }=1$%
, of course), and $\tilde{m}=m-\alpha $ is a new irrational magnetic quantum
number that indulges within the Aharonov-Bohm quantum number $\alpha =\pm
\left\vert \alpha \right\vert $ (the $\pm $ signature of $\alpha $ depends
on the positivity or negativity of the charge of the PDM-charged particle
under consideration).

Further simplification of the radial equation can be carried out by using%
\begin{equation}
R\left( \rho \right) =\sqrt{\frac{g\left( \rho \right) }{\rho }}U\left( \rho
\right) ,
\end{equation}%
to obtain the one-dimensional form of the PDM-Schr\"{o}dinger equation (14) 
\begin{equation}
\left\{ -\frac{d^{2}}{d\rho ^{2}}+\frac{\tilde{m}^{2}-1/4}{\rho ^{2}}%
+V_{eff}\left( \rho \right) +k_{z}^{2}\right\} U\left( \rho \right) =0,
\end{equation}%
where,%
\begin{equation}
V_{eff}\left( \rho \right) =V\left( \rho \right) -e\tilde{m}B_{\circ
}S\left( \rho \right) +\frac{e^{2}B_{\circ }^{2}}{4}\rho ^{2}S\left( \rho
\right) ^{2}-g\left( \rho \right) E+\left[ \frac{5}{16}\left( \frac{%
g^{\prime }\left( \rho \right) }{g\left( \rho \right) }\right) ^{2}-\frac{1}{%
4}\left( \frac{g^{\prime \prime }\left( \rho \right) }{g\left( \rho \right) }%
\right) -\frac{1}{4}\left( \frac{g^{\prime }\left( \rho \right) }{\rho
\,g\left( \rho \right) }\right) \right] .
\end{equation}%
Equation (16) is to be solved for different PDM functions and PD-magnetic
fields. Before we proceed any further, nevertheless, the contribution of
equation (13) should be made clear at this stage. As long as the
three-dimensional cylindrical settings are in point, the eigenvalues and
eigenfunctions of (13) will have their spectral signatures on the overall
spectra (on both energy eigenvalues and wave functions). Such spectral
signatures are readily and very recently discussed by Algadhi and Mustafa 
\cite{42}. The idea as well as spectral signatures are clear and need not be
repeated here again, therefore. Yet, should one be interested in the
two-dimensional flat-land polar coordinates $\left( \rho ,\varphi \right) $,
then the substitutions $Z\left( z\right) =1$, and $V\left( z\right)
=k_{z}^{2}=0$ could perfectly get the job done.

To construct the PD-magnetic fields, we observe that the choice of $S\left(
\rho \right) $, in (7), is not a random one at all. It is very much related
to the feasibly experimentally applicable nature of the PD-magnetic fields.
The choice that%
\begin{equation}
\overrightarrow{B}=B_{\circ }\left[ \frac{\mu }{\rho ^{\sigma }}\right] 
\widehat{z}\Longleftrightarrow S\left( \rho \right) +\frac{\rho }{2}%
S^{\prime }\left( \rho \right) =\frac{\mu }{\rho ^{\sigma }}%
\Longleftrightarrow S\left( \rho \right) =\left( \frac{2\mu }{2-\sigma }%
\right) \rho ^{-\sigma }+\frac{\beta }{\rho ^{2}};\,\sigma \neq 2,
\end{equation}%
looks viable and interesting. Where $\mu \neq 0$, otherwise the magnetic
field is switched off. Therefore, $S\left( \rho \right) $ works as a
generating function for the PD-magnetic fields, where for $\mu =1$ and $%
\sigma =0$ we recover the constant magnetic field settings. Nevertheless, in
the current methodical proposal we wish to work with the most simplistic
PD-magnetic field\ where $\sigma =1$, so that%
\begin{equation}
\overrightarrow{B}=B_{\circ }\left[ \frac{\mu }{\rho }\right] \,\widehat{z}%
\Longleftrightarrow S\left( \rho \right) =\frac{2\mu }{\rho }+\frac{\beta }{%
\rho ^{2}}.
\end{equation}%
This would, in turn, imply that equation (16) be rewritten as%
\begin{gather}
\left\{ -\frac{d^{2}}{d\rho ^{2}}+\frac{\tilde{m}^{2}-1/4-e\tilde{m}B_{\circ
}\beta +e^{2}B_{\circ }^{2}\beta ^{2}/4}{\rho ^{2}}-\frac{\left( 2e\tilde{m}%
B_{\circ }\mu -e^{2}B_{\circ }^{2}\mu \beta \right) }{\rho }-g\left( \rho
\right) E+V\left( \rho \right) \right.  \notag \\
\left. +\left[ \frac{5}{16}\left( \frac{g^{\prime }\left( \rho \right) }{%
g\left( \rho \right) }\right) ^{2}-\frac{1}{4}\left( \frac{g^{\prime \prime
}\left( \rho \right) }{g\left( \rho \right) }\right) -\frac{1}{4}\left( 
\frac{g^{\prime }\left( \rho \right) }{\rho \,g\left( \rho \right) }\right) %
\right] \right\} U\left( \rho \right) =\tilde{E}U\left( \rho \right) ,
\end{gather}%
where%
\begin{equation}
\tilde{E}=-\left( k_{z}^{2}+e^{2}B_{\circ }^{2}\mu ^{2}\right) .
\end{equation}%
Next, we shall be interested in a PDM in the form of%
\begin{equation}
g\left( \rho \right) =\eta \,f\left( \rho \right) \exp \left( -\delta \rho
\right)
\end{equation}%
where $f\left( \rho \right) =1$, $\delta =0$, and $\eta =1$ allow the
problem to recover constant mass settings. Yet, we shall choose some
specific values for these parameters in such a way that serves and clarifies
the current methodical proposal.

\section{Almost quasi-free, $V\left( \protect\rho \right) =0$, PDM-charged
particles in PD-magnetic and Aharonov-Bohm flux fields}

Equation (20) suggests two exactly solvable textbook-models that constitute
two \emph{almost-quasi-free} PDM-charged particles of fundamental Coulombic
nature. We use the classification \emph{almost-quasi-free} PDM-charged
particles for they are moving under the influence of only the vector
potential (8). That is, moving in the vicinity of only a PD-magnetic and an
Aharonov-Bohm flux fields (i.e. $V\left( \rho \right) =0$) . The two
examples are in order.

\subsection{An almost quasi-free PDM-charged particle of $g\left( \protect%
\rho \right) =\protect\eta /\protect\rho $}

Let us consider an almost quasi-free PDM-charged particle with $g\left( \rho
\right) =\eta /\rho $ (i.e., $f\left( \rho \right) =1/\rho $ and $\delta =0$
in (22)) moving in the vector potential (8) that yields the PD-magnetic
field of (19). Hence, equation (20) reads%
\begin{equation}
\left\{ -\frac{d^{2}}{d\rho ^{2}}+\frac{\tilde{\ell}^{2}-1/4}{\rho ^{2}}-%
\frac{\tilde{\alpha}}{\rho }\right\} U\left( \rho \right) =\tilde{E}U\left(
\rho \right) ,
\end{equation}%
where%
\begin{equation}
\tilde{\alpha}=2e\tilde{m}B_{\circ }\mu -e^{2}B_{\circ }^{2}\mu \beta +\eta
E,
\end{equation}%
and%
\begin{equation}
\tilde{\ell}^{2}=\tilde{m}^{2}+\frac{1}{16}-e\tilde{m}B_{\circ }\beta +\frac{%
e^{2}B_{\circ }^{2}\beta ^{2}}{4}\Longleftrightarrow \left\vert \tilde{\ell}%
\right\vert =\sqrt{\left( \tilde{m}-\frac{eB_{\circ }\beta }{2}\right) ^{2}+%
\frac{1}{16}}
\end{equation}%
Equation (23) is similar to the radial Schr\"{o}dinger equation of the
two-dimensional Coulombic problem and admits exact eigenvalues%
\begin{equation}
\tilde{E}=-\frac{\tilde{\alpha}^{2}}{\left[ 2\left( n_{\rho }+\left\vert 
\tilde{\ell}\right\vert +1/2\right) \right] ^{2}}\Longleftrightarrow \left(
k_{z}^{2}+e^{2}B_{\circ }^{2}\mu ^{2}\right) =\frac{\tilde{\alpha}^{2}}{%
\left[ 2\left( n_{\rho }+\left\vert \tilde{\ell}\right\vert +1/2\right) %
\right] ^{2}}\text{ };\text{ }n_{\rho }=0,1,2,\cdots .
\end{equation}%
which would in turn lead to%
\begin{equation}
E_{n_{\rho },m,\alpha }=\frac{1}{\eta }\left[ \beta \mu e^{2}B_{\circ
}^{2}-2e\left( m-\alpha \right) B_{\circ }\mu +2\sqrt{k_{z}^{2}+e^{2}B_{%
\circ }^{2}\mu ^{2}}\left( n_{\rho }+\frac{1}{2}+\sqrt{\left( m-\alpha -%
\frac{eB_{\circ }\beta }{2}\right) ^{2}+\frac{1}{16}}\right) \right] .
\end{equation}%
The radial eigenfunctions are%
\begin{equation}
R_{n_{\rho },m,\alpha }\left( \rho \right) =\mathcal{N}\rho ^{\left\vert 
\tilde{\ell}\right\vert -1/2}\exp \left( -\sqrt{k_{z}^{2}+e^{2}B_{\circ
}^{2}\mu ^{2}}\rho \right) L_{n_{\rho }}^{2\left\vert \tilde{\ell}%
\right\vert }\left( 2\sqrt{k_{z}^{2}+e^{2}B_{\circ }^{2}\mu ^{2}}\rho
\right) ,
\end{equation}%
where $L_{n_{\rho }}^{2\left\vert \tilde{\ell}\right\vert }\left( 2\sqrt{%
k_{z}^{2}+e^{2}B_{\circ }^{2}\mu ^{2}}\rho \right) $ are the Laguerre
polynomials, and $n_{\rho }$ is the radial quantum number.

\begin{figure}[t]
\includegraphics[width=70mm,scale=0.7]{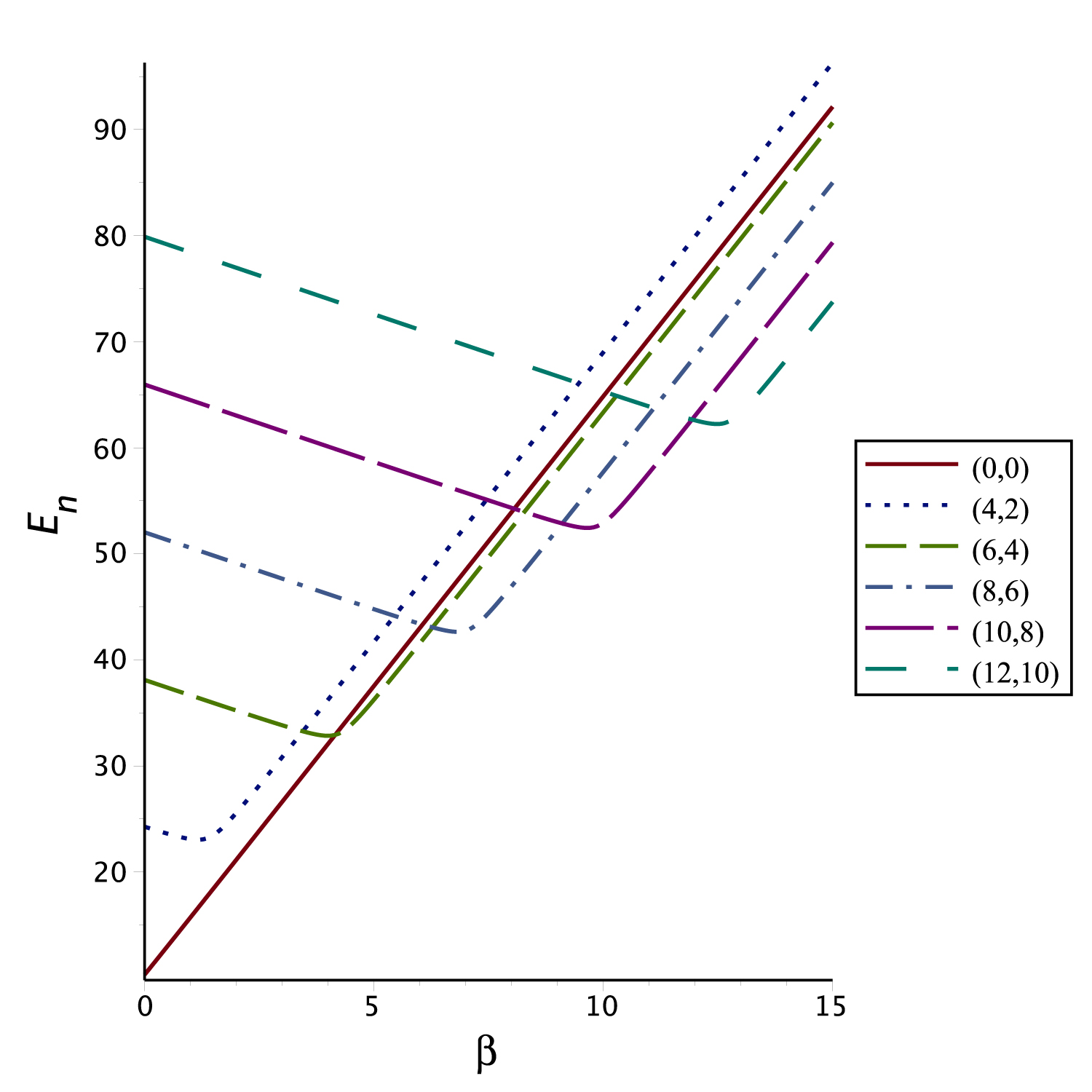}
\caption{{}Energy levels $\left( n_{\protect\rho }.m\right) $ crossings of
(27) for different values $\protect\beta $ in (19).}
\end{figure}

\begin{figure}[t]
\includegraphics[width=70mm,scale=0.7]{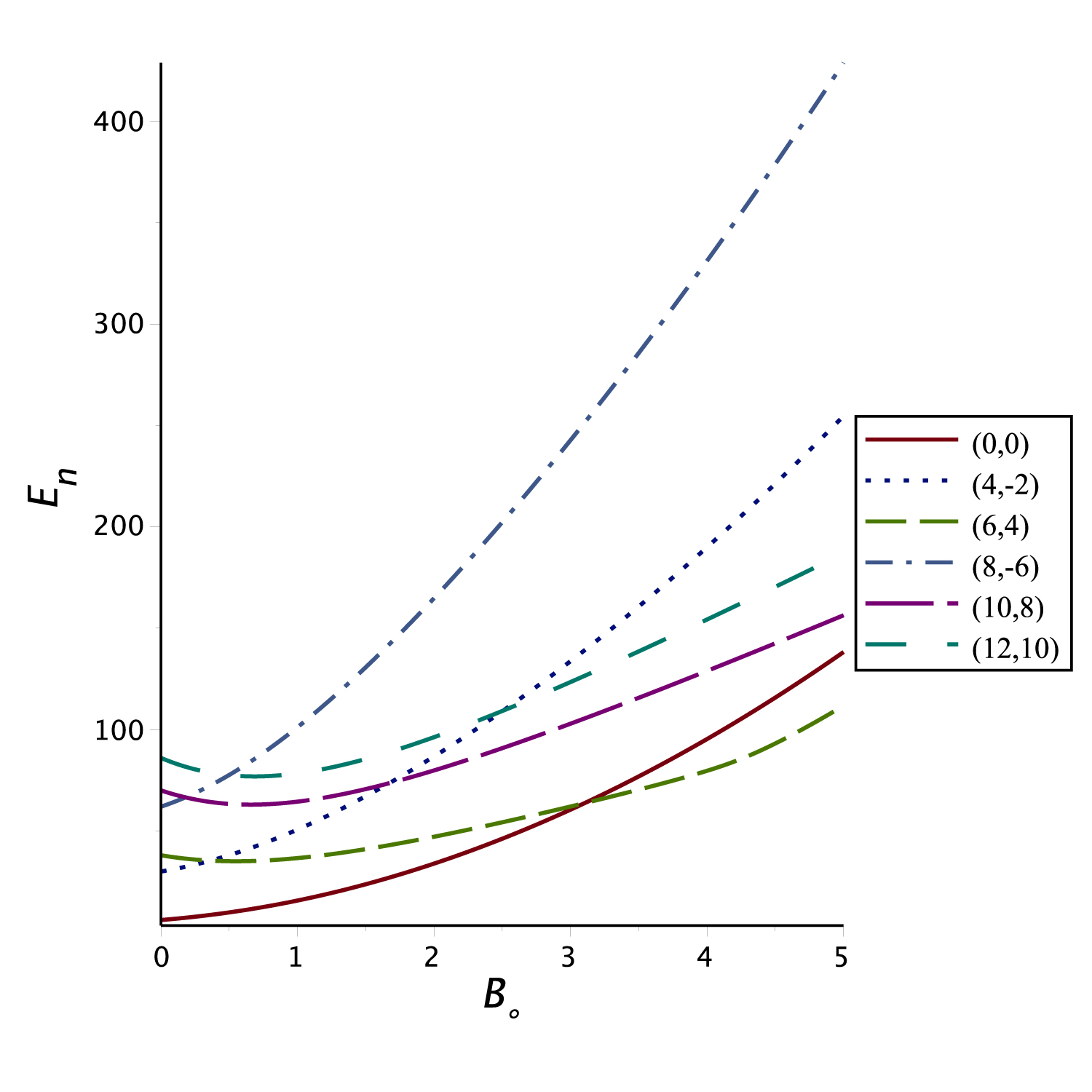}
\caption{{}Energy levels $\left( n_{\protect\rho }.m\right) $ crossings of
(27) for different values of the magnetic field strength $B_{\circ }$ in
(19).}
\end{figure}

\begin{figure}[t]
\includegraphics[width=70mm,scale=0.7]{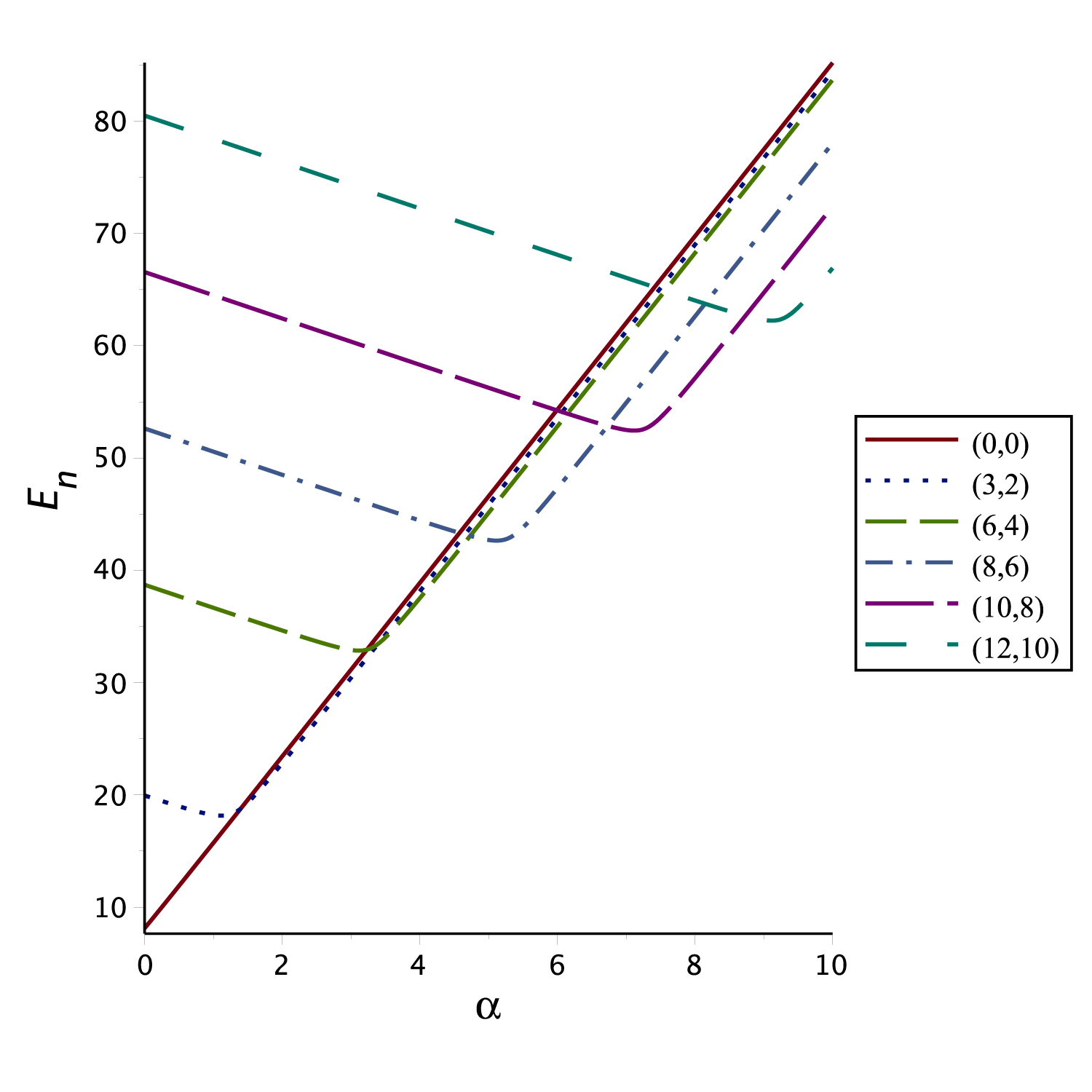}
\caption{{}Energy levels $\left( n_{\protect\rho }.m\right) $ crossings of
(27) for different Aharonov-Bohm quantum number $\protect\alpha $.}
\end{figure}

\begin{figure}[t]
\includegraphics[width=70mm,scale=0.7]{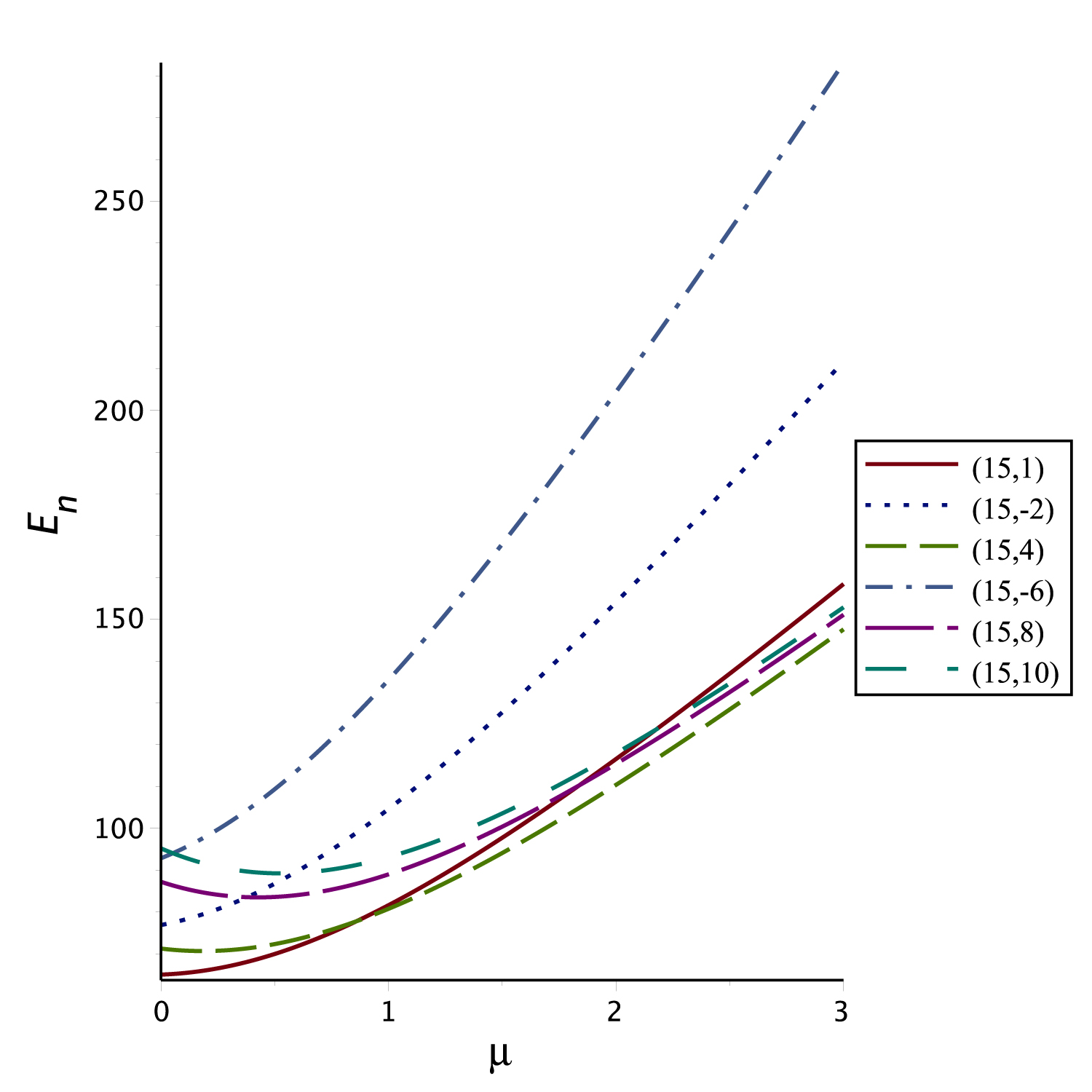}
\caption{{}Energy levels $\left( n_{\protect\rho }.m\right) $ crossings of
(27) for different values of the magnetic field parameter $\protect\mu $ in
(19).}
\end{figure}

In Fig.s 1-4 we plot the energy levels (labeled as $\left( n_{\rho
}.m\right) $) in (27) for different values of the parameters involved. The
quantum numbers of a given state $\left( n_{\rho }.m\right) $ are chosen at
random so that the phenomenon of energy levels crossings is made clear. The
energy levels crossing points suggest that there could be more than one
quantum state sharing the same energy at each crossing point. This would in
turn indicate occasional degeneracies at some specific parametric settings.

\subsection{An almost quasi-free PDM-charged particle of $g\left( \protect%
\rho \right) =\protect\eta /\protect\rho ^{2}$}

An almost quasi-free PDM-charged particle with $g\left( \rho \right) =\eta
/\rho ^{2}$ (i.e., $f\left( \rho \right) =1/\rho ^{2}$ and $\delta =0$ in
(22)) moving under the influence of\ only the vector potential (8) would
result in presenting (20) as%
\begin{equation}
\left\{ -\frac{d^{2}}{d\rho ^{2}}+\frac{\acute{\ell}^{2}-1/4}{\rho ^{2}}-%
\frac{\acute{\beta}}{\rho }\right\} U\left( \rho \right) =\tilde{E}\,U\left(
\rho \right) ,
\end{equation}%
where, 
\begin{equation}
\acute{\beta}=2e\tilde{m}B_{\circ }\mu -e^{2}B_{\circ }^{2}\mu \beta ,
\end{equation}%
and%
\begin{equation}
\acute{\ell}^{2}=\tilde{m}^{2}+\frac{1}{4}-e\tilde{m}B_{\circ }\beta +\frac{%
e^{2}B_{\circ }^{2}\beta ^{2}}{4}-\eta E\Longleftrightarrow \left\vert 
\acute{\ell}\right\vert =\sqrt{\left( \tilde{m}-\frac{eB_{\circ }\beta }{2}%
\right) ^{2}+\frac{1}{4}-\eta E}.
\end{equation}%
We have again a similar two-dimensional radial Schr\"{o}dinger equation of
Coulombic nature. One may, in a straightforward manner, show that it admits
the exact eigenvalues%
\begin{equation}
E_{n_{\rho },m,\alpha }=\frac{1}{\eta }\left[ \left( m-\alpha -\frac{%
eB_{\circ }\beta }{2}\right) ^{2}+\frac{1}{4}-\left( \frac{2e\left( m-\alpha
\right) B_{\circ }\mu -e^{2}B_{\circ }^{2}\mu \beta }{2\sqrt{%
k_{z}^{2}+e^{2}B_{\circ }^{2}\mu ^{2}}}-n_{\rho }-\frac{1}{2}\right) ^{2}%
\right] ,
\end{equation}%
and exact radial wavefunctions%
\begin{equation}
R_{n_{\rho },m,\alpha }\left( \rho \right) =\mathcal{N}\rho ^{-1+\left\vert 
\acute{\ell}\right\vert }\exp \left( -\sqrt{k_{z}^{2}+e^{2}B_{\circ }^{2}\mu
^{2}}\rho \right) L_{n_{\rho }}^{2\left\vert \acute{\ell}\right\vert }\left(
2\sqrt{k_{z}^{2}+e^{2}B_{\circ }^{2}\mu ^{2}}\rho \right)
\end{equation}

In the Fig.s 5-8 we plot the energy levels $\left( n_{\rho }.m\right) $ in
(32) for different values of the parameters involved. The the quantum states 
$\left( n_{\rho }.m\right) $ are chosen at random so that the phenomenon of
energy levels crossings is made clear. One observes multiple energy levels
crossings for each quantum state reported here. This would in turn indicate
occasional degeneracies at some specific parametric settings.

\begin{figure}[t]
\includegraphics[width=70mm,scale=0.7]{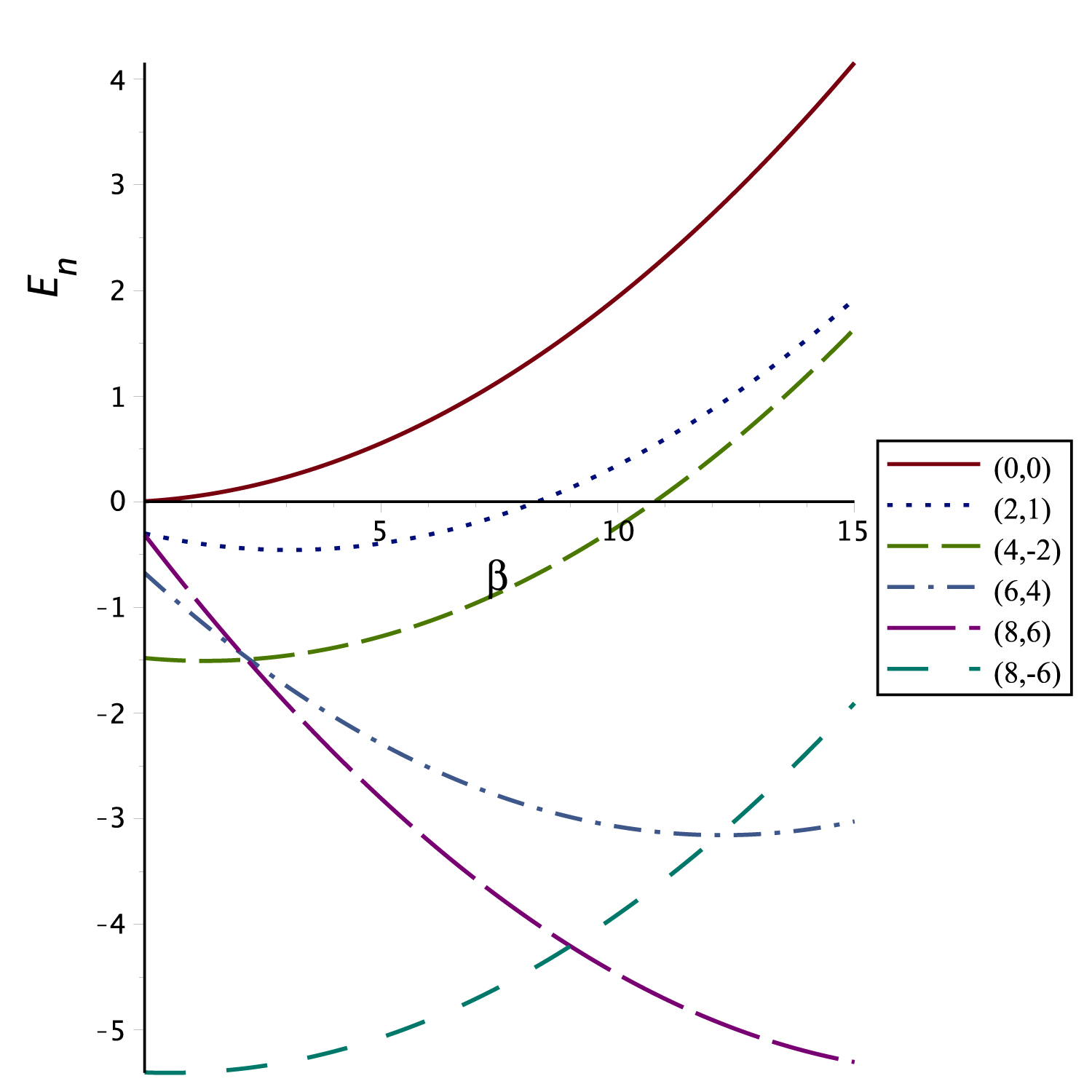}
\caption{{}Energy levels $\left( n_{\protect\rho }.m\right) $ crossings of
(32) for different values of the parameter $\protect\beta $ in (19).}
\end{figure}

\begin{figure}[t]
\includegraphics[width=70mm,scale=0.7]{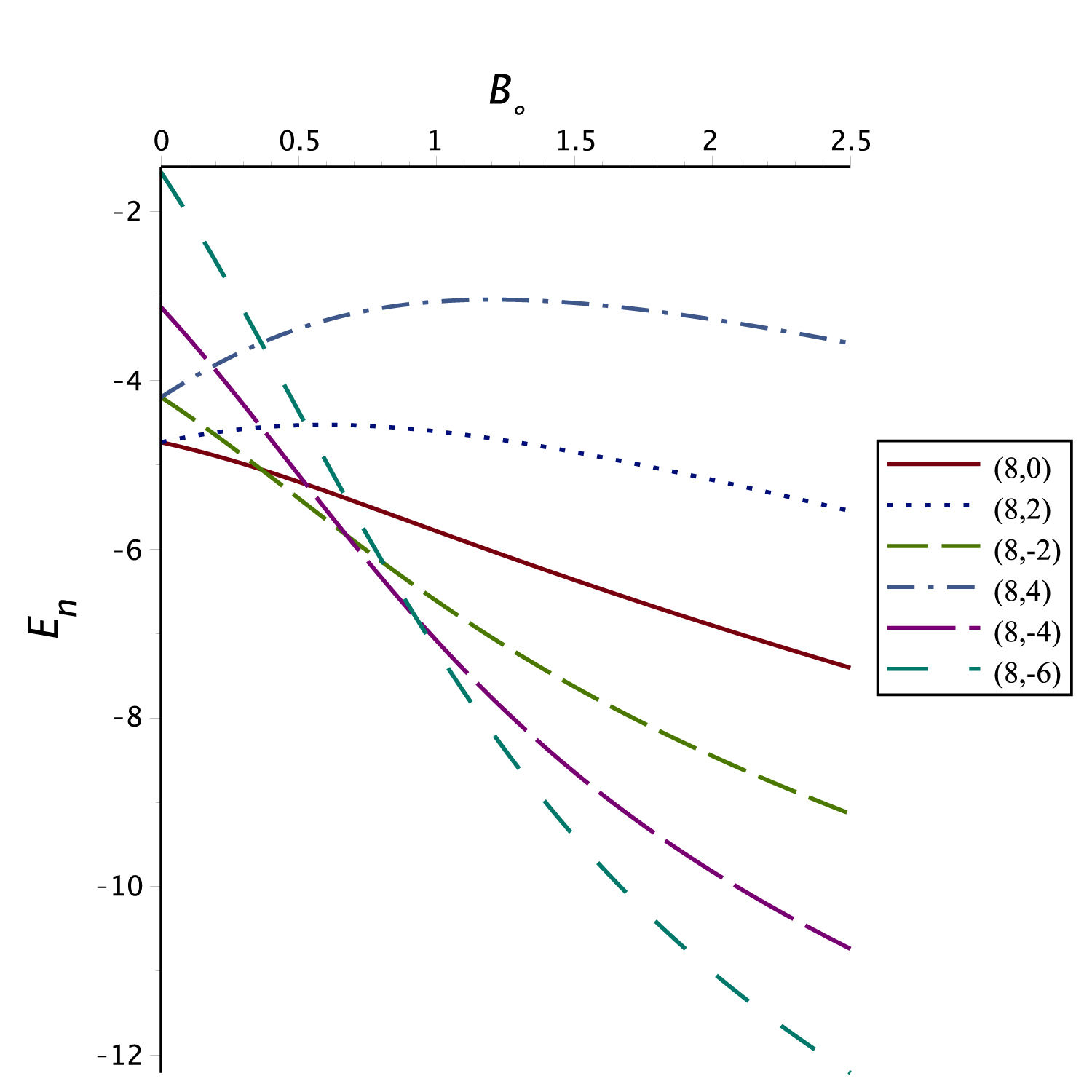}
\caption{Energy levels $\left( n_{\protect\rho }.m\right) $ crossings of
(32) for different values of the magnetic field strength $B_{\circ }$ in
(19).}
\end{figure}

\begin{figure}[t]
\includegraphics[width=70mm,scale=0.7]{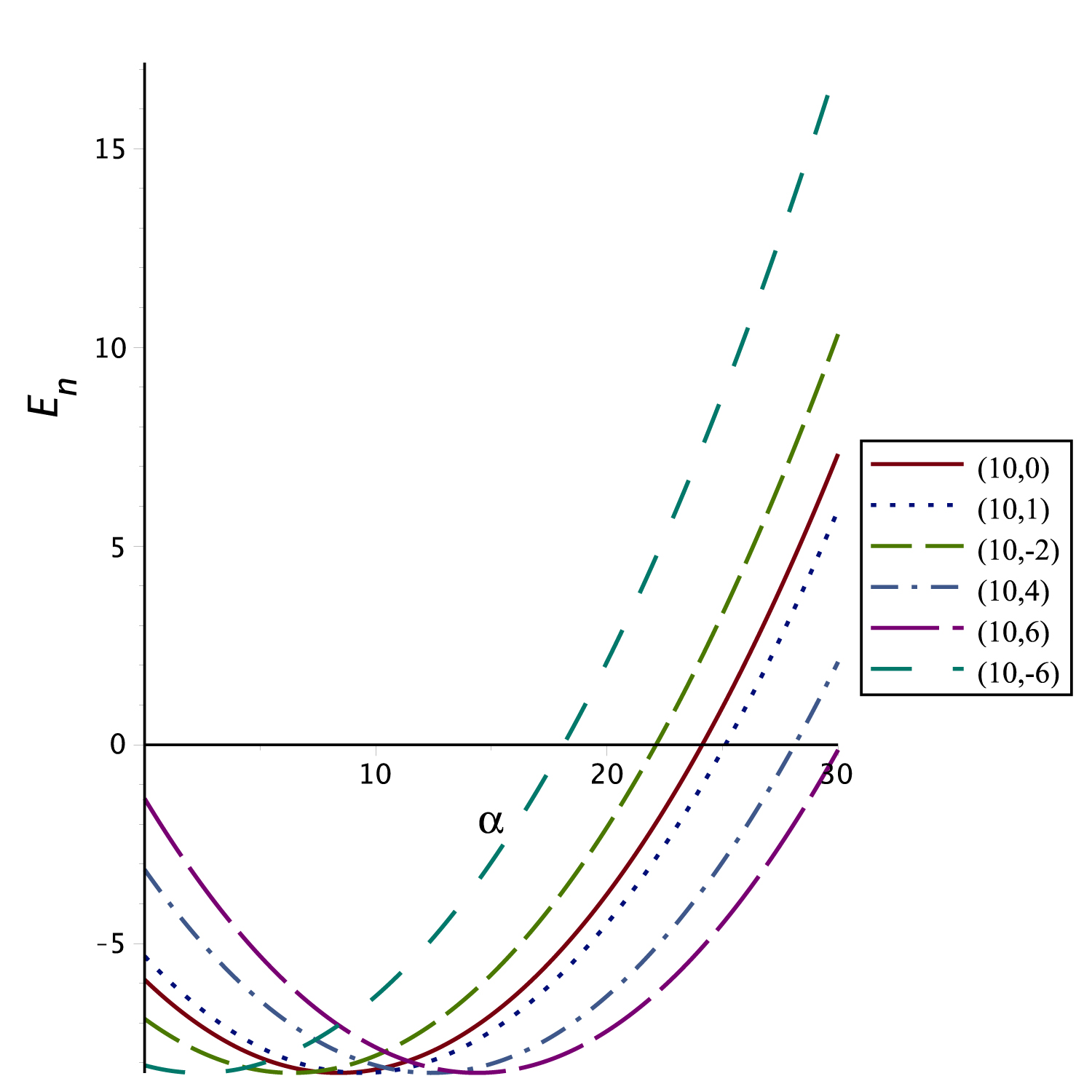}
\caption{{} Energy levels $\left( n_{\protect\rho }.m\right) $ crossings of
(32) for different values of Aharonov-Bohm quantum number$\protect\alpha $.}
\end{figure}

\begin{figure}[t]
\includegraphics[width=70mm,scale=0.7]{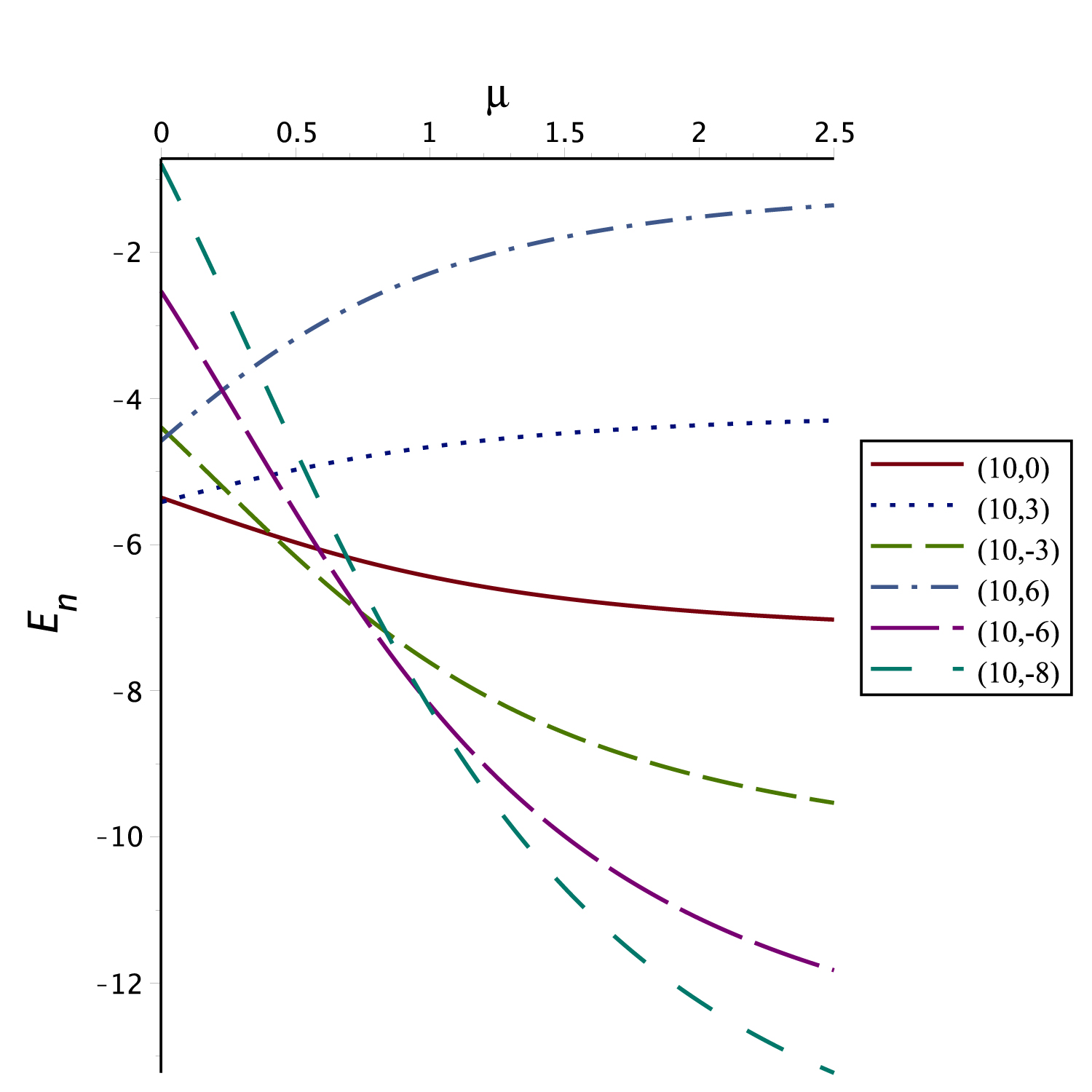}
\caption{ Energy levels $\left( n_{\protect\rho }.m\right) $ crossings of
(32) for different values of the magnetic field parameter $\protect\mu $ in
(19).}
\end{figure}

\section{PDM-charged particles in PD-magnetic and Aharonov-Bohm flux fields:
Nikiforov-Uvarov exact solvability}

In this section, we shall be interested in a PDM-charged particle endowed
with a Yukawa-type mass function%
\begin{equation}
g\left( \rho \right) =\eta \left( \frac{\exp \left( -\delta \rho \right) }{%
\rho }\right)
\end{equation}%
(i.e., $f\left( \rho \right) =1/\rho $ and $\delta \neq 0$) moving in the
vector potential (8) that yields the PD-magnetic field in (19). Moreover, we
would like to subject this PDM-charged particle to radial confining
potential of the form 
\begin{equation}
V\left( \rho \right) =-\frac{V_{\circ }\exp \left( -\delta \rho \right) }{%
\rho }-\frac{V_{_{1}}}{\rho }+\frac{V_{_{2}}}{\rho ^{2}},
\end{equation}%
which indulges within, a Yukawa-type (i.e., the first term) plus a
Kratzer-type (the last two terms) potentials. A confinement potential type
that is commonly used in the spectroscopy of the diatomic molecules, where
the Greene-Aldrich approximation%
\begin{equation}
\frac{1}{\rho }\simeq \frac{\delta }{1-\exp \left( -\delta \rho \right) }%
\Longleftrightarrow \frac{1}{\rho ^{2}}\simeq \frac{\delta ^{2}}{\left[
1-\exp \left( -\delta \rho \right) \right] ^{2}}
\end{equation}%
is valid for $\rho \ll 1$. Hence, equation (20) reads%
\begin{equation}
\left\{ -\frac{d^{2}}{d\rho ^{2}}+\frac{a_{_{1}}}{\rho ^{2}}+\frac{a_{_{2}}}{%
\rho }-a_{_{3}}\left( \frac{\exp \left( -\delta \rho \right) }{\rho }\right)
+a_{_{4}}\right\} U\left( \rho \right) =0,
\end{equation}%
where%
\begin{equation}
\begin{tabular}{cc}
$a_{_{1}}=\tilde{m}^{2}-3/16-e\tilde{m}B_{\circ }\beta +e^{2}B_{\circ
}^{2}\beta ^{2}/4+V_{_{2}}\medskip ,$ \  & $a_{_{2}}=e^{2}B_{\circ }^{2}\mu
\beta -2e\tilde{m}B_{\circ }\mu +3\delta /8-V_{_{1}}\medskip $ \\ 
$a_{_{3}}=V_{\circ }+\eta \,E\medskip $ & $a_{_{4}}=k_{z}^{2}+e^{2}B_{\circ
}^{2}\mu ^{2}+\delta ^{2}/16\medskip $%
\end{tabular}%
.
\end{equation}%
Next, the use of Greene-Aldrich approximation (36) in (37) would allow us to
rewrite it as%
\begin{equation}
\left\{ -\frac{d^{2}}{d\rho ^{2}}+\frac{a_{_{1}}\delta ^{2}}{\left[ 1-\exp
\left( -\delta \rho \right) \right] ^{2}}+\frac{a_{_{2}}\delta }{1-\exp
\left( -\delta \rho \right) }-a_{_{3}}\left( \frac{\delta \exp \left(
-\delta \rho \right) }{1-\exp \left( -\delta \rho \right) }\right)
+a_{_{4}}\right\} U\left( \rho \right) =0.
\end{equation}%
Let us now use the substitution $\xi =\exp \left( -\delta \rho \right) $ and
convert this equation into a Nikiforov-Uvarov type (see e.g. \cite{39,40,41}%
) to obtain%
\begin{equation}
\left\{ \frac{d^{2}}{d\xi ^{2}}+\frac{\left( 1-\xi \right) }{\xi \left(
1-\xi \right) }\frac{d}{d\xi }+\frac{1}{\left[ \xi \left( 1-\xi \right) %
\right] ^{2}}\left[ -\left( \tilde{a}_{_{1}}-\tilde{a}_{_{2}}+\tilde{a}%
_{_{4}}\right) +\left( -\tilde{a}_{_{2}}+\tilde{a}_{_{3}}+2\tilde{a}%
_{_{4}}\right) \,\xi -\left( \tilde{a}_{_{3}}+\tilde{a}_{_{4}}\right) \,\xi
^{2}\medskip \right] \right\} U\left( \xi \right) =0
\end{equation}%
where%
\begin{equation}
\tilde{a}_{_{1}}=a_{_{1}}\text{ , }\tilde{a}_{_{2}}=-a_{_{2}}/\delta \text{
, }\tilde{a}_{_{3}}=a_{_{3}}/\delta \text{ , }\tilde{a}_{_{4}}=a_{_{4}}/%
\delta ^{2}.
\end{equation}%
We may, therefore, express this equation in the Nikiforov-Uvarov form%
\begin{equation}
U^{\prime \prime }\left( \xi \right) +\frac{\tilde{\tau}\left( \xi \right) }{%
\sigma \left( \xi \right) }U^{\prime }\left( \xi \right) +\frac{\tilde{\sigma%
}\left( \xi \right) }{\sigma \left( \xi \right) ^{2}}U\left( \xi \right) =0,
\end{equation}%
where 
\begin{equation}
\begin{tabular}{l}
$\tilde{\tau}\left( \xi \right) \medskip =1-\xi $ , \ \ \ $\sigma \left( \xi
\right) =\xi \left( 1-\xi \right) $ \\ 
$\tilde{\sigma}\left( \xi \right) =-\left( \tilde{a}_{_{1}}-\tilde{a}_{_{2}}+%
\tilde{a}_{_{4}}\right) +\left( -\tilde{a}_{_{2}}+\tilde{a}_{_{3}}+2\tilde{a}%
_{_{4}}\right) \,\xi -\left( \tilde{a}_{_{3}}+\tilde{a}_{_{4}}\right) \,\xi
^{2}\medskip $%
\end{tabular}%
.
\end{equation}%
Which obviously satisfies the requirements of NU-method, where $\sigma
\left( \xi \right) $, $\tilde{\sigma}\left( \xi \right) $ are polynomials of
at most second degree, and $\tilde{\tau}\left( \xi \right) $ is at most a
first degree polynomial. Although the NU-method is well known, we would like
to recycle it in an optimal way that makes the current paper self-contained
and clear. We do so in the Appendix.

Following NU-method procedure of the Appendix (namely, equations (A.1) to
(A.20)), with $\tilde{a}_{_{3}}=\left( \eta E+V_{\circ }\right) /\delta $ in
(41) and (38), we obtain%
\begin{equation}
\tilde{a}_{_{3}}=\left( n_{\rho }^{2}+n_{\rho }+1/2\right) +\left( 2n_{\rho
}+1\right) \epsilon _{1}+\epsilon _{2}
\end{equation}%
where $\epsilon _{1}$ and $\epsilon _{2}$ are given through the relations $%
\epsilon _{1}=\tilde{\epsilon}_{1}/\delta $ and $\epsilon _{2}=\tilde{%
\epsilon}_{2}/\delta $ so that\ 
\begin{eqnarray}
\tilde{\epsilon}_{1} &=&\left[ \delta ^{2}\left( \tilde{m}-\frac{eB_{\circ
}\beta }{2}\right) ^{2}+\delta ^{2}V_{2}+\frac{\delta ^{2}}{4}-2eB_{\circ
}\mu \left( \tilde{m}-\frac{eB_{\circ }\beta }{2}\right) \delta -\delta
V_{1}+e^{2}B_{\circ }^{2}\mu ^{2}+k_{z}^{2}\right] ^{1/2}\medskip  \notag \\
&&+\delta \,\left[ \left( \tilde{m}-\frac{eB_{\circ }\beta }{2}\right)
^{2}+V_{2}+\frac{1}{16}\right] ^{1/2},
\end{eqnarray}%
and%
\begin{eqnarray}
\tilde{\epsilon}_{2} &=&2\left\{ \left[ \delta ^{2}\left( \tilde{m}-\frac{%
eB_{\circ }\beta }{2}\right) ^{2}+\delta ^{2}V_{2}+\frac{\delta ^{2}}{4}%
-2eB_{\circ }\mu \left( \tilde{m}-\frac{eB_{\circ }\beta }{2}\right) \delta
-\delta V_{1}+e^{2}B_{\circ }^{2}\mu ^{2}+k_{z}^{2}\right] \right.  \notag \\
&&\times \left. \left[ \left( \tilde{m}-\frac{eB_{\circ }\beta }{2}\right)
^{2}+V_{2}+\frac{1}{16}\right] \right\} ^{1/2}\medskip +2\left[ \delta
\left( \tilde{m}-\frac{eB_{\circ }\beta }{2}\right) ^{2}+\delta
V_{2}-eB_{\circ }\mu \left( \tilde{m}-\frac{eB_{\circ }\beta }{2}\right) %
\right] -V_{1}.
\end{eqnarray}%
This would eventually imply%
\begin{equation}
E_{n_{\rho },m,\alpha }=\frac{1}{\eta }\left\{ \left( n_{\rho }^{2}+n_{\rho
}+1/2\right) \delta +\left( 2n_{\rho }+1\right) \tilde{\epsilon}_{1}+\tilde{%
\epsilon}_{2}-V_{\circ }\right\}
\end{equation}%
One should notice that the result in (47) recovers that of the almost
quasi-free PDM-charged particle in (27) by setting $\delta =0$ and $V_{\circ
}=V_{1}=V_{2}=0$ in (34) and (35). This should be the typical tendency (as
well as a double check) of the exact analytical solution of the more general
problem discussed here, of course. Moreover, our radial wave functions are
given by (15), (A.1), and (A.23) to yield%
\begin{equation}
R_{n_{\rho },m,\alpha }\left( \rho \right) =\mathcal{N}_{n_{\rho }}\mathcal{%
\,\rho }^{^{-\left( 1-\upsilon \right) /2}}\exp \left( -\delta \rho \left(
1+\kappa \right) /2\right) P_{n_{\rho }}^{\left( \kappa ,\upsilon \right)
}\left( 1-2e^{-\delta \rho }\right) \medskip
\end{equation}%
where $\mathcal{N}_{n_{\rho }}$ is the corresponding normalization constant.
Yet, this result would collapse into that of the almost quasi-free
PDM-charged particle in (28) for $\delta =0$ and $V_{\circ }=V_{1}=V_{2}=0$
in (34) and (35). In Fig. 9 we plot the energies of (47) against the PDM
parameter $\delta $ of (34). We observe a direct effect of the PDM on the
energy levels crossings indicating again occasional degeneracies. 
\begin{figure}[t]
\includegraphics[width=70mm,scale=0.7]{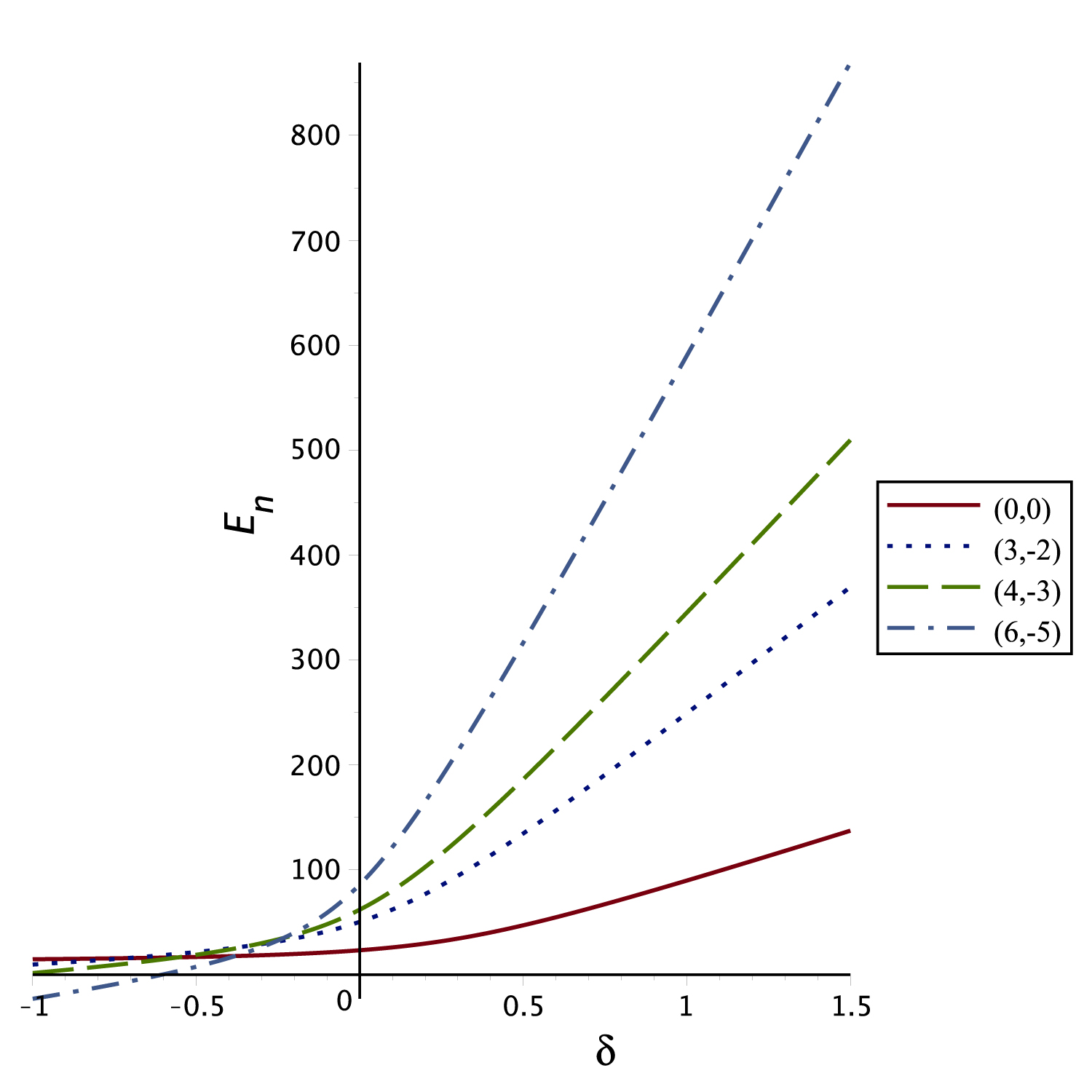}
\caption{{}Energy levels $\left( n_{\protect\rho }.m\right) $ crossings of
(47) for different values $\protect\delta $ in (34).}
\end{figure}

\section{Concluding Remarks}

We have considered, using cylindrical coordinates under azimuthal
symmetrization, some PDM-charged particles in PD-magnetic and Aharonov-Bohm
flux fields. Two \emph{almost-quasi-free} PDM-charged particles (i.e., no
conventional confinement potential, $V\left( \overrightarrow{r}\right) =0$,
and the only interaction is provided by the PDM-minimal coupling along with
the position-dependent mass) with $g\left( \rho \right) =\eta /\rho $ and $%
g\left( \rho \right) =\eta /\rho ^{2}$ turned out to imply exactly solvable
radial Schr\"{o}dinger equations of a Coulombic nature (documented in (27)
and (28) for $g\left( \rho \right) =\eta /\rho $ and (32) and (33) for $%
g\left( \rho \right) =\eta /\rho ^{2}$). Their exact solutions are inferred,
in a straightforward manner, from the textbook solutions. Moreover, a more
general Yukawa-type PDM-charged particle with $g\left( \rho \right) =\eta
\,\exp \left( -\delta \rho \right) /\rho $ moving not only in the
PD-magnetic and Aharonov-Bohm flux fields but also in the vicinity of a
Yukawa plus a Kratzer type confinement potential field $V\left( \rho \right)
=-V_{\circ }\exp \left( -\delta \rho \right) /\rho -V_{_{1}}/\rho
+V_{_{2}}/\rho ^{2}$ is considered. For this case, we have used the
NU-method to obtain exact analytical eigenvalues and eigenfunctions
(reported in (47) and (48), respectively). Our observations are in order.

The energy levels crossings are observed eminent for the two \emph{%
almost-quasi-free} PDM-charged particles in (27) ( for $g\left( \rho \right)
=\eta /\rho $) and (32) (for $g\left( \rho \right) =\eta /\rho ^{2}$). We
have, therefore, reported Figures 1-8 so that this phenomenon of levels
crossings is made clear. The quantum states involved in the plots are
labeled as $\left( n_{\rho }.m\right) $ and the quantum numbers $n_{\rho }$
and $m$ are chosen at random. Such energy levels crossing points suggest
that there could be more than one quantum state sharing the same energy at
each crossing point. For example, Fig.1 shows the effect of the parameter $%
\beta $ of the magnetic field generating function (19) on the energy levels
of (27). Hereby, we observe that the state labeled $\left( 12,10\right) $
crosses with $\left( 4,2\right) $, $\left( 6,4\right) $, $\left( 8,6\right) $%
, $\left( 0,1\right) $, $\left( 10,8\right) $. This would indicate that the
state $\left( 12,10\right) $ share the same energy with $\left( 4,2\right) $%
, $\left( 6,4\right) $, $\left( 8,6\right) $, $\left( 0,1\right) $, $\left(
10,8\right) $ at the crossing points. Similar scenarios of energy levels
crossings are also observed in Fig.s 2-8. Where, Fig.s 1 and 5 show the
effect of $\beta $, Fig.s 2 and 6 show the effect of the magnetic field
strength $B_{\circ }$ of (19), Fig.s 3 and 7 show the effect the
Aharonov-Bohm quantum number $\alpha $, and Fig.s 4 and 8 show the effect
the magnetic field parameter $\mu $ on the energy levels of (27) and (32),
respectively. Such energy levels crossings may very well be classified as "%
\emph{occasional degeneracies}" that have erupted as a result of PDM
setting. Energy levels crossings are also observed feasible in (47) for the
Yukawa-type PDM-charged particles of (34) confined in the
Yukawa-plus-Kratzer potential (35). This is documented in Fig. 9, where a
direct influence of the PDM-parameter $\delta $ of (34) on the energy levels
crossings is observed.

On the position-dependent settings side, we have chosen, in (19), to work
with a PD-magnetic field $\overrightarrow{B}=\left( B_{\circ }\mu /\rho
\,\right) \widehat{z}$ (i.e., $\sigma =1$ in (18)). In so doing, we sought
simplicity and physical eligibility in order to make the current methodical
proposal instructive \ and clear. Of course, one may wish to work with other
values for $\sigma $ in (18), as yet another form for the PD-magnetic field,
and follow the same procedure to find the eigenvalues and eigenfunctions.
Moreover, our choices for the PDM-functions in (23), (29), and (34) are
manifested meanly by \ the convenience of the current study.

Finally, we have asserted that a deformation on the coordinate system may
very well render the mass position-dependent \cite{11,31} (i.e., the mass
becomes metaphorically speaking position-dependent). One may, therefore, use
the eigen energies to calculate the partition functions and discuss some
thermodynamical properties (see, e.g., \cite{41,44,45}) of such PDM systems
in PD-magnetic and Aharonov Bohm fields and, perhaps, in diatomic confining
potentials like the Tietz oscillator, Rosen-Morse, Manning-Rosen, etc.

\section{Appendix: Nikiforov-Uvarov method}

In this section, we recycle it in an optimal way that makes the current
paper self-contained and clear. We, therefore, closely follow Badalov \cite%
{39} (where instructive and informative details on NU-method are available).
As such, a substitution of the form%
\begin{equation}
U\left( \xi \right) =\phi \left( \xi \right) \chi \left( \xi \right) 
\tag*{(A.1)}
\end{equation}%
in (42) would lead to a hypergeometric-type equation%
\begin{equation}
\sigma \left( \xi \right) \chi ^{\prime \prime }\left( \xi \right) +\tau
\left( \xi \right) \chi ^{\prime }\left( \xi \right) +\lambda \chi \left(
\xi \right) =0  \tag*{(A.2)}
\end{equation}%
where its solutions $\chi _{n}\left( \xi \right) $ satisfy the Rodrigues
relation%
\begin{equation}
\chi _{n}\left( \xi \right) =\frac{A_{n}}{\omega \left( \xi \right) }\frac{%
d^{n}}{d\xi ^{n}}\left( \sigma ^{n}\left( \xi \right) \omega \left( \xi
\right) \right) \text{ };\text{ \ }n=0,1,2,\cdots  \tag*{(A.3)}
\end{equation}%
and the weight function $\omega \left( \xi \right) $ satisfies the condition%
\begin{equation}
\frac{d}{d\xi }\left( \sigma \left( \xi \right) \omega \left( \xi \right)
\right) =\tau \left( \xi \right) \omega \left( \xi \right) .  \tag*{(A.4)}
\end{equation}%
Here, we have used 
\begin{equation}
\pi \left( \xi \right) =\sigma \left( \xi \right) \frac{\phi ^{\prime
}\left( \xi \right) }{\phi \left( \xi \right) }\text{, }\tau \left( \xi
\right) =\tilde{\tau}\left( \xi \right) +2\pi \left( \xi \right) \text{ , \ }%
\lambda =\frac{\phi ^{\prime \prime }\left( \xi \right) }{\phi \left( \xi
\right) }\sigma \left( \xi \right) +\frac{\tilde{\tau}\left( \xi \right) }{%
\sigma \left( \xi \right) }\pi \left( \xi \right) +\frac{\tilde{\sigma}%
\left( \xi \right) }{\sigma \left( \xi \right) }  \tag*{(A.5)}
\end{equation}%
with the condition $\tau ^{\prime }\left( \xi \right) <0$ is imposed on the
weight function $\omega \left( \xi \right) $. The parameter $\lambda $
required for this method is\ also defined as%
\begin{equation}
\lambda =k+\pi ^{\prime }\left( \xi \right) .  \tag*{(A.6)}
\end{equation}%
Hence, (A.5) and (A.6) yield%
\begin{equation}
\pi \left( \xi \right) =\frac{\sigma ^{\prime }\left( \xi \right) -\tilde{%
\tau}\left( \xi \right) }{2}\pm \sqrt{\frac{\left( \sigma ^{\prime }\left(
\xi \right) -\tilde{\tau}\left( \xi \right) \right) ^{2}}{4}+k\sigma \left(
\xi \right) -\tilde{\sigma}\left( \xi \right) }.  \tag*{(A.7)}
\end{equation}

To find the value of $k$ in (A.7), one should be able to write the
expression under the square root in the form of a quadratic equation (i.e.,
square of a polynomial of first degree). That is, the expression under the
square root should look like $A\xi ^{2}+B\xi +C\Longleftrightarrow \left( 
\sqrt{A}\xi \pm \sqrt{C}\right) ^{2}$ so that $B=\pm 2\sqrt{AC}$ is a
condition imposed on the adjustable parameter $k$. Moreover, the eigenvalues
of the hypergeometric equation (A.2) are given by%
\begin{equation}
\lambda =\lambda _{n}=-n\tau ^{\prime }\left( \xi \right) -\frac{n\left(
n-1\right) }{2}\sigma ^{\prime \prime }\left( \xi \right) \text{ };\text{ }%
n=0,1,2,\cdots  \tag*{(A.8)}
\end{equation}%
Consequently, our $\pi \left( \xi \right) $ is given by%
\begin{equation}
\pi \left( \xi \right) =-\frac{\xi }{2}\pm \sqrt{A\xi ^{2}+B\xi +C}=-\frac{%
\xi }{2}\pm \left( \sqrt{A}\xi \pm \sqrt{C}\right) ;\,B=\pm 2\sqrt{AC}, 
\tag*{(A.9)}
\end{equation}%
where%
\begin{equation}
\begin{tabular}{lll}
$A=1/4-k+\tilde{a}_{_{3}}+\tilde{a}_{_{4}},$ & $B=k+\tilde{a}_{_{2}}-\tilde{a%
}_{_{3}}-2\tilde{a}_{_{4}},$ & $C=\tilde{a}_{_{1}}-\tilde{a}_{_{2}}+\tilde{a}%
_{_{4}}$%
\end{tabular}
\tag*{(A.10)}
\end{equation}%
At this point, we recollect that%
\begin{equation}
\tau \left( \xi \right) =\tilde{\tau}\left( \xi \right) +2\pi \left( \xi
\right) \Longleftrightarrow \tau ^{\prime }\left( \xi \right) =-2\pm 2\sqrt{A%
},  \tag*{(A.11)}
\end{equation}%
which suggests that the condition $\tau ^{\prime }\left( \xi \right) <0$ is
satisfied if and only if 
\begin{equation}
\tau ^{\prime }\left( \xi \right) =-2-2\sqrt{A}\Longleftrightarrow \pi
\left( \xi \right) =\pi _{_{-}}\left( \xi \right) =-\frac{\xi }{2}-\left( 
\sqrt{A}\xi \pm \sqrt{C}\right) \Leftrightarrow \tau \left( \xi \right) =%
\tilde{\tau}\left( \xi \right) +2\pi _{_{-}}\left( \xi \right) , 
\tag*{(A.12)}
\end{equation}%
provided that $A>0$ in (A.10) otherwise unphysical imaginary energy
eigenvalues are obtained by (A.8). Moreover, the relation $B=\pm 2\sqrt{AC}%
\Longleftrightarrow B^{2}=4AC$ in (A.9) would imply that 
\begin{equation*}
k=k_{_{\pm }}=-\left( 2\tilde{a}_{_{1}}-\tilde{a}_{_{2}}-\tilde{a}%
_{_{3}}\right) \pm \sqrt{\left( \tilde{a}_{_{1}}-\tilde{a}_{_{2}}+\tilde{a}%
_{_{4}}\right) \left( 4\tilde{a}_{_{1}}+1\right) }.
\end{equation*}%
Consequently, the condition that $A=1/4-k+\tilde{a}_{_{3}}+\tilde{a}%
_{_{4}}>0 $ of (A.10) would necessarily imply that 
\begin{equation}
k=k_{_{-}}=-\left( 2\tilde{a}_{_{1}}-\tilde{a}_{_{2}}-\tilde{a}%
_{_{3}}\right) -\sqrt{\left( \tilde{a}_{_{1}}-\tilde{a}_{_{2}}+\tilde{a}%
_{_{4}}\right) \left( 4\tilde{a}_{_{1}}+1\right) }.  \tag*{(A.13)}
\end{equation}%
Now we go back to our $\pi \left( \xi \right) $ of (A.9), along with (A.10)
and $k=k_{_{-}}$ in (A.13), and cast it as 
\begin{equation}
\pi \left( \xi \right) =\pi _{_{-}}\left( \xi \right) =-\frac{\xi }{2}-\sqrt{%
A_{_{-}}\xi ^{2}+B_{_{-}}\xi +C},  \tag*{(A.14)}
\end{equation}%
where, in this case, equation (A.10) suggests that 
\begin{equation}
A=A_{_{-}}=\frac{1}{4}-k_{_{-}}+\tilde{a}_{_{3}}+\tilde{a}_{_{4}}=\left( 
\frac{4\tilde{a}_{_{1}}+1}{4}\right) +\left( \tilde{a}_{_{1}}-\tilde{a}%
_{_{2}}+\tilde{a}_{_{4}}\right) +\sqrt{\left( \tilde{a}_{_{1}}-\tilde{a}%
_{_{2}}+\tilde{a}_{_{4}}\right) \left( 4\tilde{a}_{_{1}}+1\right) }, 
\tag*{(A.15)}
\end{equation}%
and%
\begin{equation}
B=B_{_{-}}=k_{_{-}}+\tilde{a}_{_{2}}-\tilde{a}_{_{3}}-2\tilde{a}_{_{4}}=-2%
\sqrt{\tilde{a}_{_{1}}-\tilde{a}_{_{2}}+\tilde{a}_{_{4}}}\left( \sqrt{\tilde{%
a}_{_{1}}-\tilde{a}_{_{2}}+\tilde{a}_{_{4}}}+\sqrt{\frac{4\tilde{a}_{_{1}}+1%
}{4}}\right) .  \tag*{(A.16)}
\end{equation}%
Next, a straightforward rearrangement of the terms in (A.14) immediately
yields%
\begin{equation}
\pi _{_{-}}\left( \xi \right) =-\frac{\xi }{2}-\left[ \left( \sqrt{\tilde{a}%
_{_{1}}-\tilde{a}_{_{2}}+\tilde{a}_{_{4}}}+\sqrt{\frac{4\tilde{a}_{_{1}}+1}{4%
}}\right) \xi -\sqrt{\tilde{a}_{_{1}}-\tilde{a}_{_{2}}+\tilde{a}_{_{4}}}%
\right]  \tag*{(A.17)}
\end{equation}%
This would, in turn, imply that the condition 
\begin{equation}
\tau ^{\prime }\left( \xi \right) =-2-2\left( \sqrt{\tilde{a}_{_{1}}-\tilde{a%
}_{_{2}}+\tilde{a}_{_{4}}}+\sqrt{\frac{4\tilde{a}_{_{1}}+1}{4}}\right) <0. 
\tag*{(A.18)}
\end{equation}%
is satisfied. We are now at a point where we can calculate the eigenvalues
working with (A.6)%
\begin{equation}
\lambda =k_{_{-}}+\pi _{_{-}}^{\prime }\left( \xi \right) =-\left( 2\tilde{a}%
_{_{1}}-\tilde{a}_{_{2}}-\tilde{a}_{_{3}}\right) -\sqrt{\left( \tilde{a}%
_{_{1}}-\tilde{a}_{_{2}}+\tilde{a}_{_{4}}\right) \left( 4\tilde{a}%
_{_{1}}+1\right) }-\frac{1}{2}-\sqrt{\tilde{a}_{_{1}}-\tilde{a}_{_{2}}+%
\tilde{a}_{_{4}}}-\sqrt{\frac{4\tilde{a}_{_{1}}+1}{4}}  \tag*{(A.19)}
\end{equation}%
and (A.8)%
\begin{equation}
\lambda =\lambda _{n}=-n\tau ^{\prime }\left( \xi \right) -\frac{n\left(
n-1\right) }{2}\sigma ^{\prime \prime }\left( \xi \right) =n\left[ 2+2\left( 
\sqrt{\tilde{a}_{_{1}}-\tilde{a}_{_{2}}+\tilde{a}_{_{4}}}+\sqrt{\frac{4%
\tilde{a}_{_{1}}+1}{4}}\right) \right] +n\left( n-1\right)  \tag{(A.20)}
\end{equation}

On the other hand, the eigenfunctions $U\left( \xi \right) =\phi \left( \xi
\right) \chi \left( \xi \right) $ are obtained in a straightforward manner.
That is,%
\begin{equation}
\frac{\phi ^{\prime }\left( \xi \right) }{\phi \left( \xi \right) }=\frac{%
\pi _{_{-}}\left( \xi \right) }{\sigma \left( \xi \right) }%
\Longleftrightarrow \phi \left( \xi \right) =\xi ^{\sqrt{\tilde{a}_{_{1}}-%
\tilde{a}_{_{2}}+\tilde{a}_{_{4}}}}\left( 1-\xi \right) ^{\frac{1}{2}+\frac{1%
}{2}\sqrt{4\tilde{a}_{_{1}}+1}},\medskip  \tag*{(A.21)}
\end{equation}%
and the weight function $\omega \left( \xi \right) $ is calculated through
(A.4) to obtain 
\begin{equation}
\frac{\omega ^{\prime }\left( \xi \right) }{\omega \left( \xi \right) }=%
\frac{\tau \left( \xi \right) -\sigma ^{\prime }\left( \xi \right) }{\sigma
\left( \xi \right) }\Longleftrightarrow \omega \left( \xi \right) =\xi ^{2%
\sqrt{\tilde{a}_{_{1}}-\tilde{a}_{_{2}}+\tilde{a}_{_{4}}}}\left( 1-\xi
\right) ^{\sqrt{4\tilde{a}_{_{1}}+1}}.  \tag*{(A.22)}
\end{equation}%
Consequently, the Rodrigues relation (A.3), with $\kappa =2\left( \sqrt{%
\tilde{a}_{_{1}}-\tilde{a}_{_{2}}+\tilde{a}_{_{4}}}\right) $ and $\upsilon =%
\sqrt{4\tilde{a}_{_{1}}+1}$, implies%
\begin{equation}
\chi _{n_{\rho }}\left( \xi \right) =A_{n_{\rho }}\,\xi ^{-\kappa }\left(
1-\xi \right) ^{-\upsilon }\frac{d^{n_{\rho }}}{d\xi ^{n_{\rho }}}\left[ \xi
^{n_{\rho }+\kappa }\left( 1-\xi \right) ^{n_{\rho }+\upsilon }\right] =%
\tilde{A}_{n_{\rho }}\,P_{n_{\rho }}^{\left( \kappa ,\upsilon \right)
}\left( 1-2\xi \right) \,;\text{ \ }\xi \in \left[ 0,1\right] \medskip 
\tag*{(A.23)}
\end{equation}%
where $P_{n}^{\left( \kappa ,\upsilon \right) }\left( 1-2\xi \right) $ are
the Jacobi polynomials satisfying the Rodrigues' formula%
\begin{equation}
P_{n}^{\left( \alpha ,\beta \right) }\left( x\right) =\frac{\left( -1\right)
^{n}}{2^{n}n!}\left( 1-x\right) ^{-\alpha }\left( 1+x\right) ^{-\beta }\frac{%
d^{n}}{dx^{n}}\left[ \left( 1-x\right) ^{n+\alpha }\left( 1+x\right)
^{n+\beta }\right] \,;\,x\in \left[ -1,1\right] .\medskip  \tag*{(A.24)}
\end{equation}%
Hereby, we have used $x=2\xi -1$ along with the polynomials symmetry
relation 
\begin{equation}
P_{n}^{\left( \alpha ,\beta \right) }\left( x\right) =\left( -1\right)
^{n}P_{n}^{\left( \beta ,\alpha \right) }\left( -x\right)
\Longleftrightarrow P_{n}^{\left( \upsilon ,\kappa \right) }\left( 2\xi
-1\right) =\left( -1\right) ^{n}P_{n}^{\left( \kappa ,\upsilon \right)
}\left( 1-2\xi \right) ,  \tag*{(A.25)}
\end{equation}%
( with $\beta =\kappa $ and $\alpha =\upsilon $) to obtain (A.23).

\end{document}